\documentclass[conference]{IEEEtran}
\usepackage{amsmath,graphicx}
\usepackage{xcolor}
\usepackage[caption=false]{subfig}
\usepackage{enumitem}
\usepackage{textcomp}
\PassOptionsToPackage{normalem}{ulem}
\usepackage{ulem}
\usepackage[unicode=true, bookmarks=false, breaklinks=false,pdfborder={0 0 1},backref=false,colorlinks=false]{hyperref}
\usepackage{flushend}
\usepackage{amssymb}
\usepackage{url}
\usepackage{tabularx} 
\usepackage{booktabs} 
\usepackage{array}
\usepackage{multirow,rotating,makecell}
\usepackage{cleveref}

\newcommand{\matlab}{MATLAB\textsuperscript{\textregistered}}
\IEEEoverridecommandlockouts
\begin{document}

\title{Fristograms: Revealing and Exploiting Light Field Internals}

\author{\IEEEauthorblockN{Thorsten Herfet, Kelvin Chelli, Tobias Lange and Robin Kremer\thanks{The work underlying this paper has been funded by the German National Science Foundation (DFG) under the project FiDaLiS, grant number 429078454.}}
\IEEEauthorblockA{Saarland Informatics Campus\\D-66123 Saarbr\"ucken, Germany\\Email: {\{herfet, chelli, lange, kremer\}@cs.uni-saarland.de}}}

\maketitle

\begin{abstract}
In recent years, light field (LF) capture and processing has become an integral part of visual media production. Several kinds of LF displays have been realized as well. The richness of information available in LFs has enabled novel applications in art, science, engineering, and medicine. Some of those in the area of computer vision are post-capture depth-of-field editing, 3D reconstruction, segmentation and matting, saliency detection, object detection and recognition, and mixed reality. The efficacy of such applications depends on certain underlying requirements, which are often ignored. For example, some operations such as noise-reduction, shift-sum- or hyperfan-filtering are only possible if a scene point, when seen from different perspectives, doesn't significantly change its appearance (Lambertian radiation). Some other operations such as the removal of obstacles or looking behind objects are only possible if there is at least one ray capturing the required scene point. Consequently, the ray distribution representing a certain scene point is an important characteristic for evaluating processing possibilities. Having access to a large number of rays allows for e.g. noise reduction, fewer rays allow for realistic reproduction of partially occluded scene elements. In this paper, we propose a novel framework for the analysis and processing of LFs that relates the captured rays to the cameras in the underlying camera setup.  

The primary idea in this paper is to establish a relation between the underlying capturing setup and the rays of the LF. To this end, we discretize the view frustum. Traditionally, a uniform discretization of the view frustum results in voxels that represents a single sample, or data point, on a regularly spaced, three-dimensional grid. Instead, we use frustum-shaped voxels, called froxels, by using depth and capturing-setup-dependent discretization of the view frustum. Froxels have been proposed earlier for novel volumetric rendering applications in modern games and 3D displays. Based on such discretization, we count the number of rays mapping to the same pixel or group of pixels on the capturing device(s). By means of this count, we propose histograms of ray-counts over the froxels (\emph{fristograms}). Fristograms can be used as a tool to analyze and reveal interesting aspects of the underlying LF, like the number of rays originating from a scene point and the color distribution of these rays. As an example, we show its ability by significantly reducing the number of rays which enables noise reduction while maintaining the realistic rendering of non-Lambertian or partially occluded regions.

\end{abstract}

\begin{IEEEkeywords}
light fields, froxels, light field analysis, froxel histograms
\end{IEEEkeywords}

\IEEEpeerreviewmaketitle

\section{Introduction\label{sec:Introduction}}

While it took more than a decade for the light field (LF) capture to gain popularity after the introduction of the first single-lens plenoptic camera, the LF media ecosystem has seen a rapid progress in the recent years~\cite{lfhistory,6277403,debevec}. In plenoptic cameras for the professional market~\cite{raytrix,klens}, as well as in mobile phones with assemblies of multiple cameras~\cite{iphone12,pixel5,nokia9},
 the underlying theory of LF processing~\cite{levoy1996light} is the foundation of depth-image-based rendering (DIBR) for applications such as super-resolution~\cite{le2020high}, de-noising~\cite{alain2017light} and depth-dependent blurring e.g. for portraits. LF processing offers a cornucopia of different enhancements. Removal of obstacles is possible if at least one subaperture view (SAI) captures the scene points occluded by the obstacle, looking around corners (optical parallax) is possible when the background can be correctly inpainted, noise reduction can be done by averaging several rays stemming from the same scene point and refocussing can be done by shift-sum- or hyperfan-filtering~\cite{dansereau2015linear}. 
 
 Recently, significant progress has been achieved by means of new LF representations such as Fourier Disparity Layers~\cite{le2019fourier}, and multi-plane- resp. multi-sphere- images, enabling real-time multi-view rendering~\cite{flynn2019deepview,broxton2020deepview}. Common to most of these techniques is the explicit or implicit generation of depth/disparity information. Both, Fourier Disparity Layers and Multi-Plane-Images detect depth layers with a significant signal energy, i.e. those layers that contain objects. Since for many scenes the number of required depth layers is limited, this can lead to a more efficient representation and a low complexity view interpolation. The semantics of the underlying assembly of rays, however, is not exploited or even gets lost~\cite{wu2017light}.
 
We emphasize the importance of the set of rays captured by a LF and used for processing due to the following reasons:
\begin{enumerate}[leftmargin=*]
\item LF capture generates huge amounts of data. The LF array built in our lab generates \textasciitilde 70~Gbps\footnote{1920 $\times$ 1200 $\times$ 40 $\times$ 12 $\times$ 64~bps}~\cite{herfet2019}. 

\item Most algorithms superimposing (in the simplest cases averaging) different rays assume Lambertian radiation. Consequently artifacts in refocussed or denoised LF images are mostly visible in regions with non-Lambertian properties.

\item No representation currently available has an explicit notion of \emph{time}. Especially in regions of very fast motion it might be advantageous to generate different views over time rather than over space. 

\end{enumerate}

In this paper, we propose a novel framework for the analysis and processing of LFs that relates the captured rays to the cameras in the underlying camera setup and allows for the \emph{semantic} analysis of the captured set of rays and application of the most appropriate signal processing. The framework and the underlying theory is described in \Cref{sec:CamArray}. We also show a simple downsampling of the rays for noise reduction to validate the applicability. The \matlab source code used to generate the fristograms and the scene reconstruction is made available as open source software\footnote{Code repository available at: \url{https://git.nt.uni-saarland.de/FiDALiS/SemanticAnalysisofLightfields}}.

\section{Theoretical background}
\label{sec:CamArray}
We now elaborate the theoretical underpinnings of this paper. We begin with a detailed description of froxels (\Cref{ssec:froxels}). Next, we explain Fristograms and the method to generate them (\Cref{subsec:Fristograms}), followed by example applications of Fristograms (\Cref{subsec:Fristogram-Applications})
\subsection{Froxels}
\label{ssec:froxels}
\begin{figure}
    \centering
    \subfloat[Uniform discretization: Voxels]{\includegraphics[width=0.75\columnwidth]{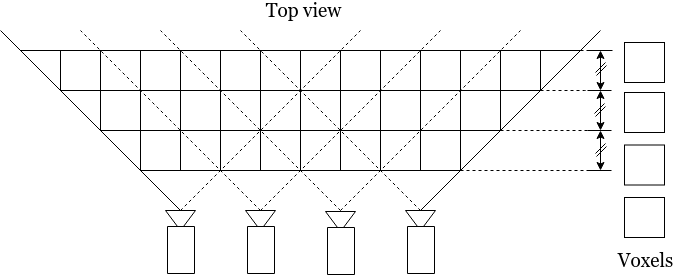}}\\
    \subfloat[Non-uniform discretization: Froxels]{\includegraphics[width=0.75\columnwidth]{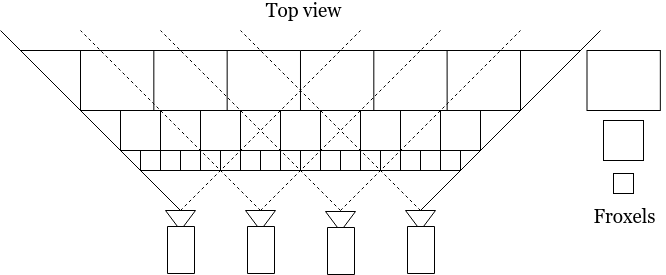}}
    \caption{Discretization of the view frustum.}
    \label{fig:froxel-voxel}
\end{figure}

The major difference between LF and point cloud capture lies in the fact that in LF capture, several scene points are captured multiple times. Thus, the number of occurrences of the same scene point in the captured set of rays is an important parameter. With the limited resolution of capturing devices --- each pixel captures a single ray --- and assuming knowledge of the camera parameters, we can define a 3-dimensional subspace whose shape and size follows the frustum. Such a 3-dimensional subspace is called a \emph{froxel}. The term froxel first appeared in~\cite{FroxelTerm2} and can be defined as a voxel that follows the shape of the frustum~\cite{vasilakis2020survey,slabaugh2000volumetric,gkioxari2019mesh}. As shown in \Cref{fig:froxel-voxel}, a uniform discretization of the view frustum results in voxels. On the other hand, a depth- and capturing-setup-dependent discretization results in froxels.

In this paper, we propose the froxel to have an extent of one~pixel in each direction. In the camera plane, this describes the frustum of a single pixel, while the extent in the depth is derived by a disparity change of one pixel. Accordingly, the width $w_{froxel}$ and height $h_{froxel}$ are proportional to (i) the pixel size $p$ on the camera sensor (assuming rectangular pixels), and (ii) the distance of the froxel from the camera plane $D_{plane}$, and inversely proportional to the focal distance of the camera $f_{d}$. Equation~\eqref{eq:FroxelSize} is used to compute the width and the height of a froxel. When choosing the scaling factors $n_{hor}=1$ and $n_{ver}=1$, the 2D cross section of a froxel is exactly the size of one camera pixel at the distance $D_{plane}$. Froxels can also be configured to have multiple pixels in the horizontal and vertical dimension by increasing $n_{hor}$ and $n_{ver}$ in~\eqref{eq:FroxelSize}. 

\begin{equation}
\begin{aligned}w_{froxel} & =\frac{n_{hor}p}{f_{d}}D_{plane},\;\;\;h_{froxel}=\frac{n_{ver}p}{f_{d}}D_{plane}.\end{aligned}
\label{eq:FroxelSize}
\end{equation}

Next, we define the depth of the froxels. When generating a LF, different view points are captured by a camera array. The frustums of the individual cameras overlap and a disparity between the cameras can be calculated. We define the depth of one froxel to be equal to the depth covered by one pixel disparity to the neighboring camera. This way, the frustum is also discretized along the camera rays and gives the froxels a depth. For a given distance $D_{plane}$, the depth per pixel disparity can be computed using~\eqref{eq:depthperpixelDisp}. 

\begin{equation}
d_{froxel}=\left.{D_{plane}^{2}}\middle/\left({\frac{f_{d}\cdot b\cdot(N-1)}{p}-D_{plane}}\right)\right..
\label{eq:depthperpixelDisp}
\end{equation}

Here, $b$ is the camera baseline. We notice that the width and height of a froxel scale linearly with the distance from the camera array while the depth of a froxel scales quadratically. As an example with the set of parameters of our own physical camera array~\cite{herfet2019}, if we consider the distance of a scene point to the camera array $D_{plane}=$ 2000mm, focal distance $f_{d}=$ 12.5mm, camera baseline $b=$ 70mm and the size of a pixel on the sensor to be $p=$ 5.86$\mu$m, the width and height of the froxel at depth $D_{plane}$ is $w_{froxel}=h_{froxel}=$ 0.938mm $\simeq$ 1mm for $n_{hor} = n_{ver}=1$. The depth of the froxel is calculated to be $d_{froxel}=$ 27.15mm.

Thus, the underlying idea is to discretize the scene into a set of froxels as described in this section. Once the froxels are defined, we assign all rays of the LF to the froxel from which they originate. This representation allows to derive simple measures like ray count but also more complex characteristics like the color distribution (to judge noise or Lambertian-ness). It also allows us to visualize the ray distribution in a scene using Fristograms (discussed in \Cref{subsec:Fristograms}) and leverage this information for LF processing. 


\begin{figure}[t!]
\centering{}
\includegraphics[width=\columnwidth]{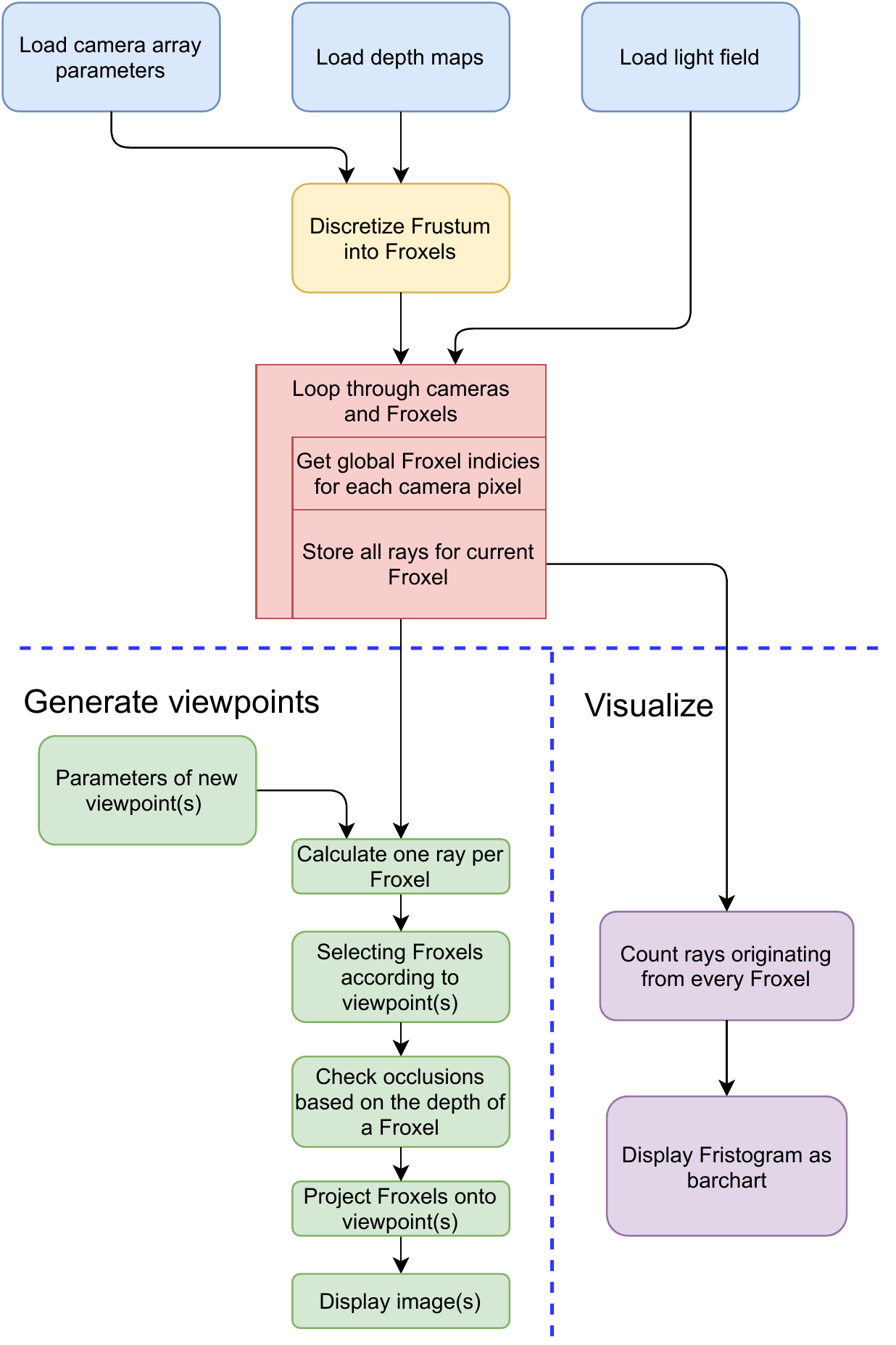}
\caption{Workflow of Fristograms generation.}
\label{fig:pipeline}
\end{figure}

\subsection{Semantic Analysis Framework}
\label{subsec:Fristograms}

An important prerequisite to semantically analyze LFs using froxels are depth maps. This requirement is not unusual as several standardization activities\footnote{JPEG PLENO: https://jpeg.org/jpegpleno/}\textsuperscript{,}\footnote{\url{MPEG-MIV: https://mpeg.chiariglione.org/sites/default/files/files/standards/parts/docs/w19212.zip}} make extensive use of depth resp. disparity maps to eliminate the redundancy stemming from Lambertian scene points captured by several cameras while not sacrificing the quality of scene points visible for only few or even one camera. JPEG Pleno~\cite{astola2020jpeg} uses disparity maps to detect occlusions and disocclusions while MPEG favors the generation of so-called atlases~\cite{boyce2021mpeg}. Thus, the depth/disparity maps --- whether measured/estimated or generated synthetically --- are a common requirement for most of the standard LF processing steps. In this work, we use depth maps to discretize the view frustum into froxels.

Depth maps for the chosen scenes were generated using a Blender addon library which is implemented based on the works of Honauer~\textit{et al.}~\cite{honauer2016benchmark}. The addon creates a virtual configurable camera array which can be placed at the desired position in the scene. Then, a color image from every camera is rendered using the chosen rendering engine. After rendering the colour images, the renderer switches to an OpenGL-based rendering engine and activates a configurable oversampling factor which increases the rendering resolution, re-renders the scene and then stores the depth and disparity information in the \emph{pfm} format with both the original and increased resolution.

The proposed framework to semantically analyze LFs and visualize ray distribution in the form of Fristograms is shown in \Cref{fig:pipeline}. The depth maps of the scene and the parameters of the camera array are used to discretize the frustum of the camera array.
Once the froxels are defined, all the rays of the underlying LF are assigned to the froxels from which they originate. This provides an insight into the distribution of rays i.e. how densely a given froxel is sampled, how many cameras capture a scene point within that froxel, the origin of each ray and the color distribution of the rays in a given froxel. This data can also be used to visualize the ray distribution in the form of Fristograms and be leveraged to perform semantically meaningful LF processing. 

Moreover, new view points of the scene can also be generated using the framework. For this, we collect all froxels that lie within the frustum of the new view point. The next step is to take occlusions into account by selecting the closest non-empty froxels to the camera. This is possible since the proposed semantics also gives us the distance of every froxel to the camera plane. Finally, the froxels are projected onto a camera that lies at the location of the new view point.

\begin{figure}[t]
\centering{}
\subfloat[Non-Lambertian]{\includegraphics[width=0.3\columnwidth]{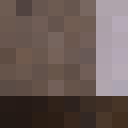}}\,\,
\subfloat[Lambertian]{\includegraphics[width=0.3\columnwidth]{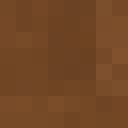}}
\caption{Examples of color distribution in froxels for certain scene points.}
\label{fig:LambertianAnalysis}
\end{figure}
\begin{figure}[t]
\centering{}
\subfloat[Gaussian noise\label{fig:NoiseAnalysisGaussian}]{\includegraphics[width=0.3\columnwidth]{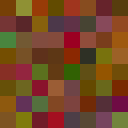}}\,\,
\subfloat[Salt-and-pepper noise\label{fig:NoiseAnalysisSaltAndPepper}]{\includegraphics[width=0.3\columnwidth]{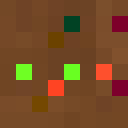}}
\caption{Froxel data with noise.}
\label{fig:NoiseAnalysis}
\end{figure}
\begin{figure}
    \centering
    \includegraphics[width=\columnwidth]{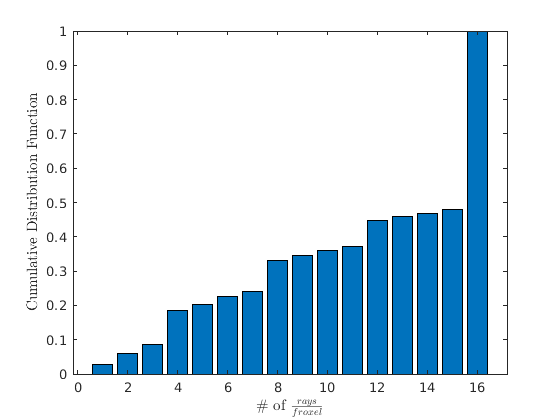}
    \caption{CDF of the ray count for the 'Classroom' scene (empty froxels omitted).}
    \label{fig:cdf}
\end{figure}
\begin{figure*}[t]
    \centering
    \setlength{\tabcolsep}{2pt}
    \begin{tabular}{cccc}
    &BMW       & Classroom & Wanderer\\
    {\begin{sideways}\,\,\,\,\,\,\,\,Original\end{sideways}}&
    \subfloat{\includegraphics[width=0.2\textwidth]{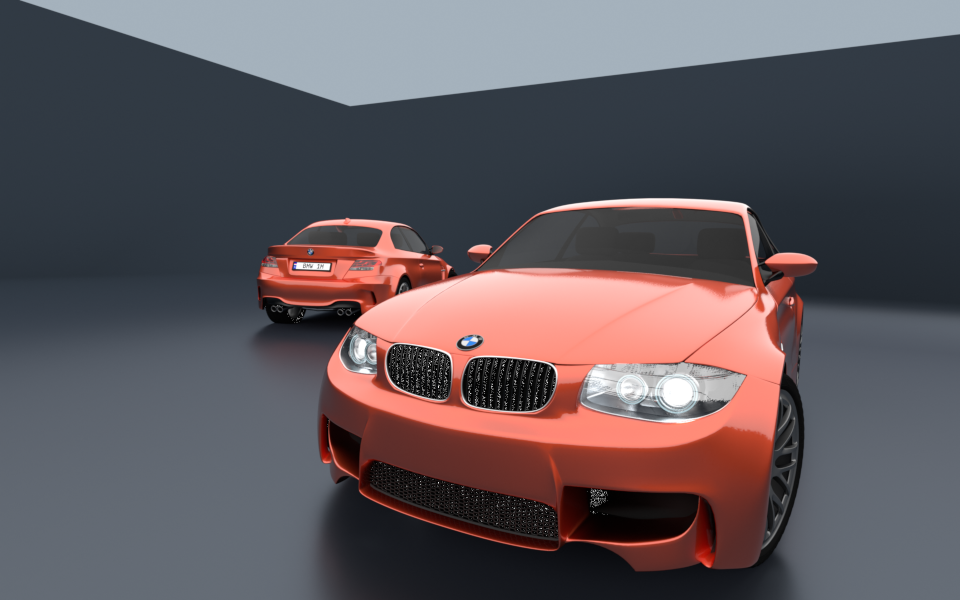}}&
    \subfloat{\includegraphics[width=0.2\textwidth]{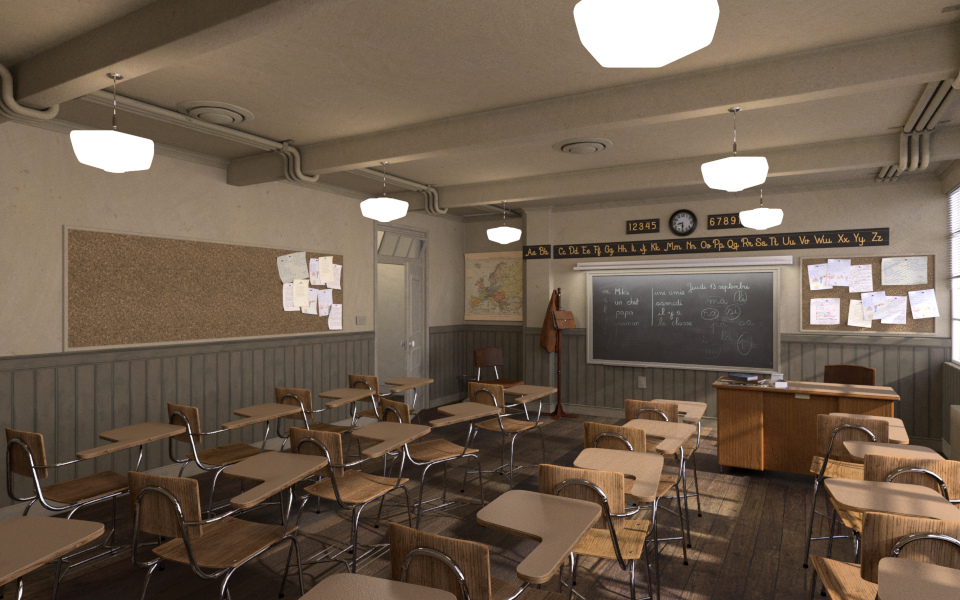}}&
    \subfloat{\includegraphics[width=0.2\textwidth]{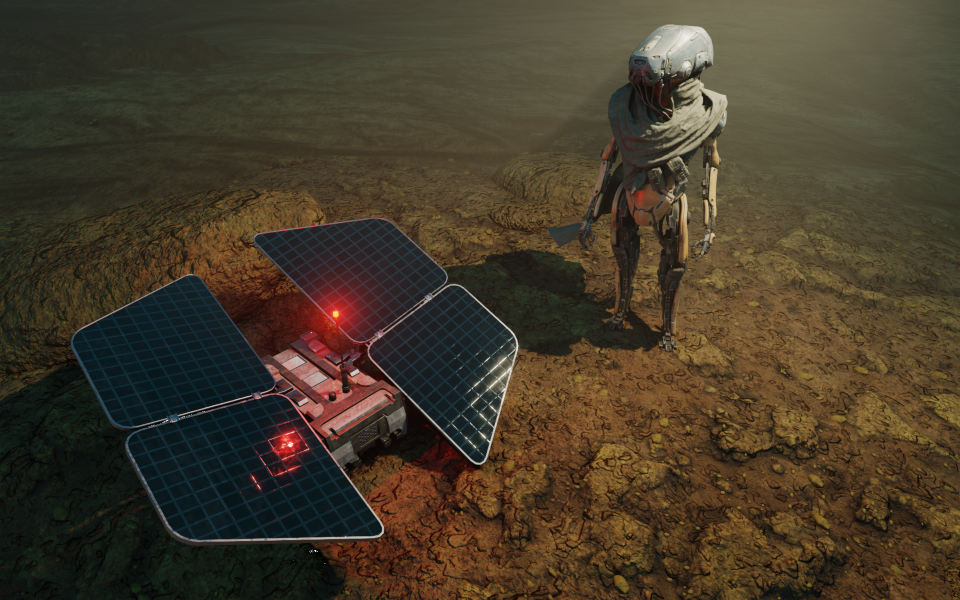}}\\
    {\begin{sideways}\,\,\,\,\,Synthesized\end{sideways}}&
        \subfloat{\includegraphics[width=0.2\textwidth]{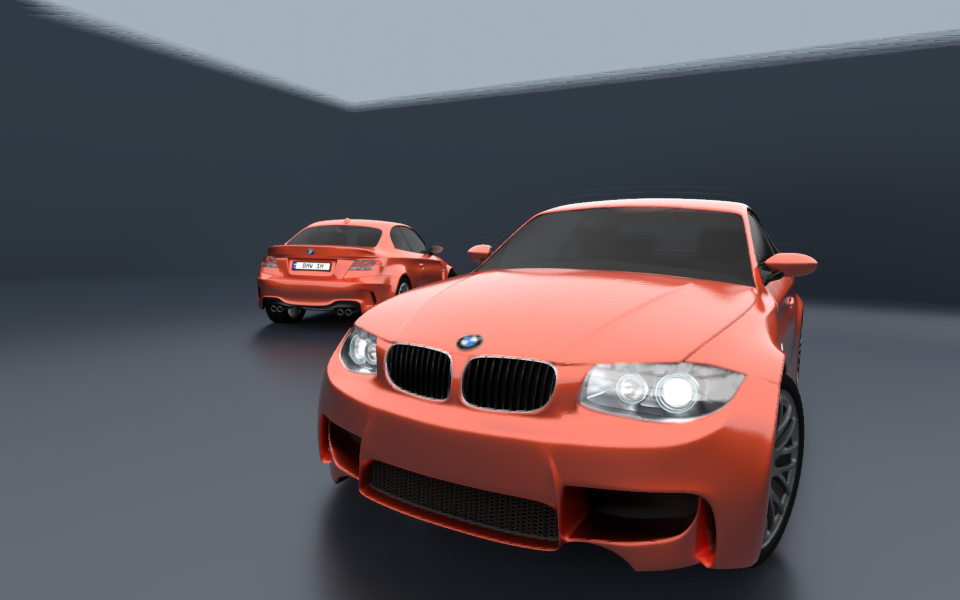}}&
    \subfloat{\includegraphics[width=0.2\textwidth]{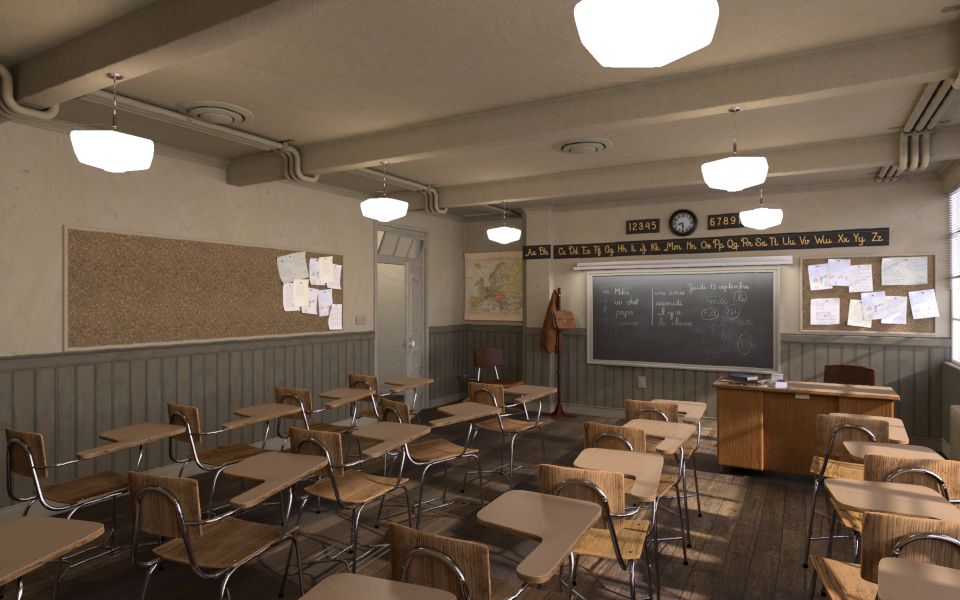}}&
    \subfloat{\includegraphics[width=0.2\textwidth]{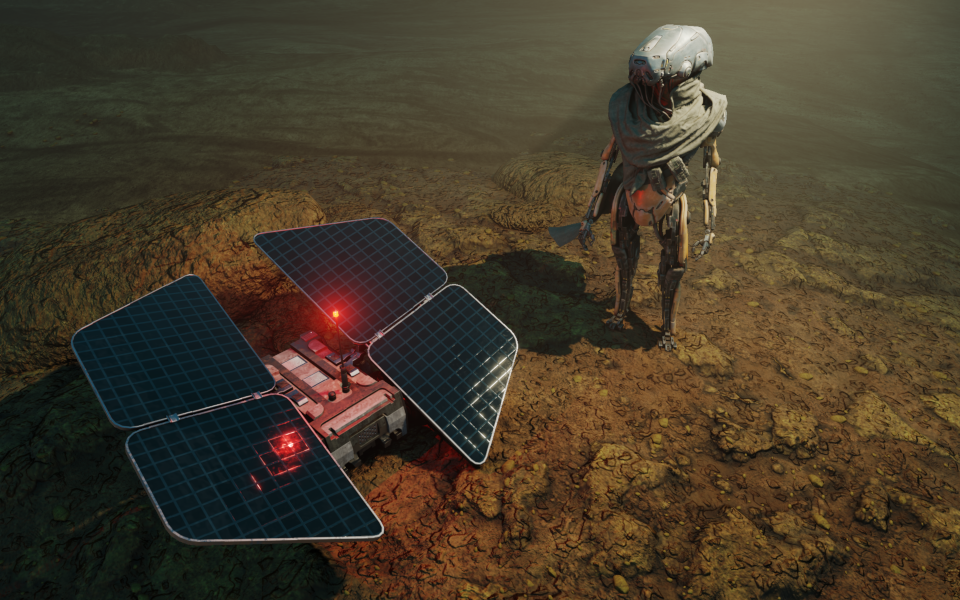}}\\
    &SSIM = 0.9684       & SSIM = 0.9658 & SSIM = 0.9833\\

    \end{tabular}
    \caption{LF at viewpoint (a, b) = (2, 2).}
    \label{fig:OrigVsReconst}
\end{figure*}
\begin{table*}[ht]
\caption{Results of semantic analysis and denoising.}
\centering
\resizebox{0.7\textwidth}{!}{
\begin{tabular}{ccccccccc}\toprule
\multirow{2}{*}{\makecell{LF\\Asset}} & \multirow{2}{*}{\makecell{Original\\size (MB)}} & \multirow{2}{*}{\makecell{Reduced\\size 
(MB)}} & \multirow{2}{*}{\makecell{Noise\\type}} & \multicolumn{2}{c}{SSIM}&&\multicolumn{2}{c}{PSNR (dB)} \\ \cmidrule{5-6}\cmidrule{8-9}
 &  &  &  & Noisy LF  & Denoised LF&&Noisy LF  & Denoised LF \\ \midrule
\multirow{2}{*}{BMW} & \multirow{2}{*}{184}  & \multirow{2}{*}{16} & Gaussian &0.3242  & 0.7678&& 20.28 & 27.43 \\\cmidrule{4-6}\cmidrule{8-9}
 &  &  & Salt-and-pepper & 0.2975 & 0.9647&& 17.96 & 29.61 \\\midrule
\multirow{2}{*}{Classroom} & \multirow{2}{*}{184} & \multirow{2}{*}{16} & Gaussian & 0.4606 & 0.8644&& 20.22 & 28.89 \\\cmidrule{4-6}\cmidrule{8-9}
 &  &  & Salt-and-pepper & 0.4136 & 0.9609&& 18.18 & 32.31 \\\midrule
\multirow{2}{*}{Wanderer} & \multirow{2}{*}{184} & \multirow{2}{*}{12} & Gaussian & 0.4350 & 0.8656&& 20.48 & 30.33\\\cmidrule{4-6}\cmidrule{8-9}
 &  &  & Salt-and-pepper & 0.3781 & 0.9821&& 17.64 & 34.78\\\bottomrule
\end{tabular}}
\label{tab:results}
\end{table*}

\subsection{Fristograms Applications\label{subsec:Fristogram-Applications}}

Fristograms provide a flexible tool to analyze and process LFs. The exact number of rays in each froxel gives us an indication of potential post-processing capabilities. We can also infer the origins of the rays and their respective color information precisely. This information is valuable to analyze the Lambertian nature of the underlying scene. We can also exploit this information to eliminate potential outliers that might e.g. stem from imperfect depth maps. In \Cref{fig:LambertianAnalysis}, we show the color distributions of 64 rays, from an 8 $\times$ 8 camera array, that originate in two different froxels from the `Classroom' scene available publicly as Blender demo files~\cite{blenderdemofiles}.
The left image shows a non-Lambertian scene point (shiny metal part of a chair with specular reflection). The variations in the color values, stemming from the different reflections seen by different cameras, are evident and can easily be detected by simple algorithms. The right image shows a Lambertian radiator (from the teachers wooden table), where we can see a near uniform distribution of color, as expected. Note that, although the examples shown in this paper depict colour distribution in froxels containing 64 rays arranged as 8 $\times$ 8 patches, the number of rays in every froxel can vary between 0 and the number of sub-aperture views M in the capturing setup. 

Fristograms also enable us to analyze the effect of noise on the rays in a froxel. \Cref{fig:NoiseAnalysis} compares Gaussian and salt-and-pepper noise on the rays. In the context of froxels, the Gaussian noise represents heavy camera noise in low-light situations and the salt-and-pepper noise stands for possibly dead pixels in the cameras. In \Cref{fig:NoiseAnalysisGaussian}, Gaussian noise is simulated with mean $\mu=$ 0 and normalized variance $\sigma^{2}=$ 0.01 while in \Cref{fig:NoiseAnalysisSaltAndPepper}, salt-and-pepper noise with 5$\%$ noise density is simulated. Our global vision is to use Fristograms as tool to analyze 5D LFs (each ray has four spatial and one temporal dimension) and derive spatio-temporal sampling patterns optimized to the intricacies of the scene in addition to developing novel filtering and post processing algorithms.

\begin{figure*}[t]
    \centering
    \setlength{\tabcolsep}{2pt}
    \begin{tabular}{cccc|cc}
    && LF with added  & Synthesized LF       & LF with salt-and-    & Synthesized LF\\
    && Gaussian noise & after mean filtering & pepper noise         & after median filtering\\
    {\begin{sideways}{\hspace{0.8cm}BMW}\end{sideways}}&&
    \subfloat[SSIM = 0.3242\label{fig:GaussNoisyBMW}]{\includegraphics[width=0.45\columnwidth]{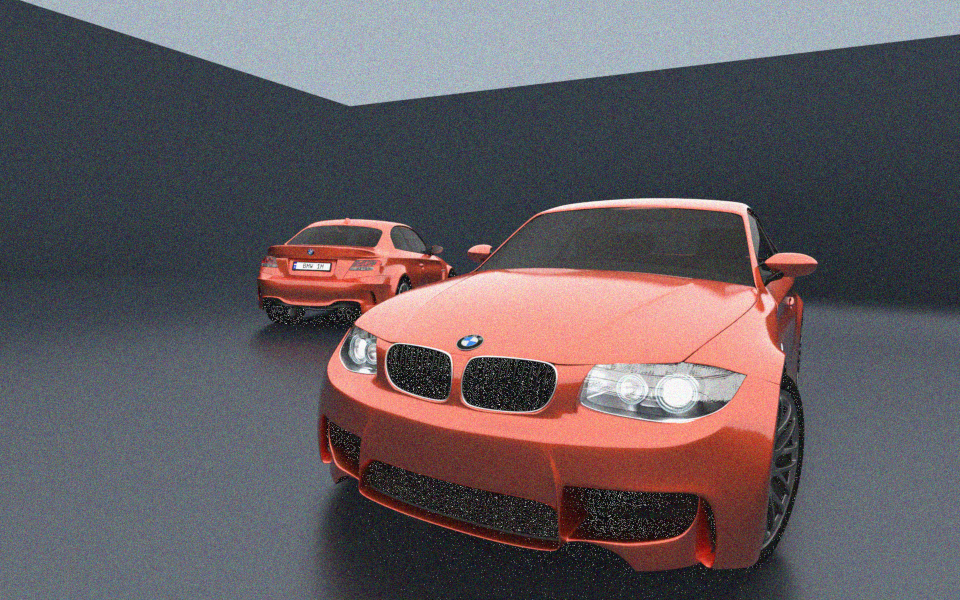}}&
    \subfloat[SSIM = 0.7678\label{fig:GaussReconstBMW}]{\includegraphics[width=0.45\columnwidth]{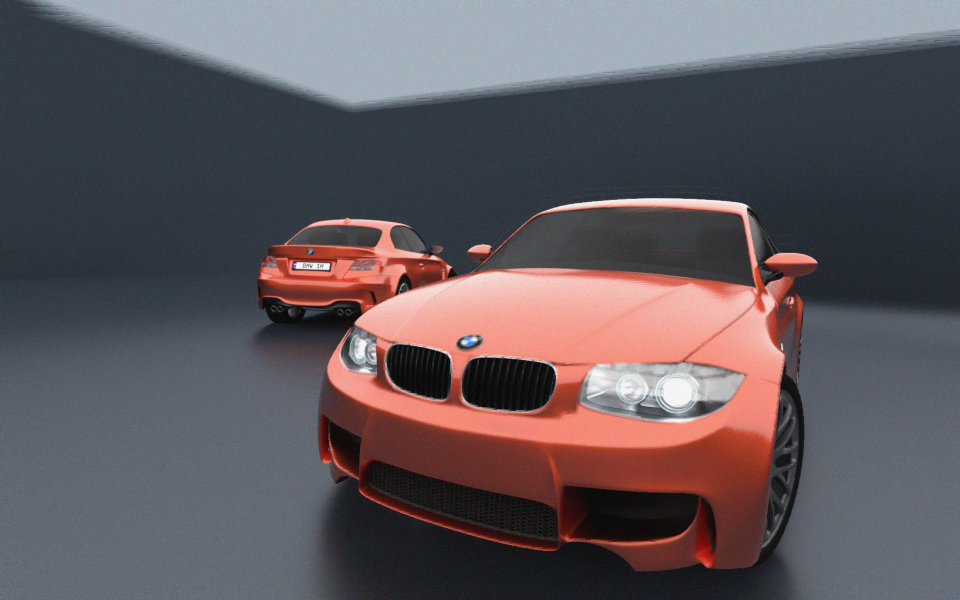}}&
    \subfloat[SSIM = 0.2975\label{fig:SalzNoisyBMW}]{\includegraphics[width=0.45\columnwidth]{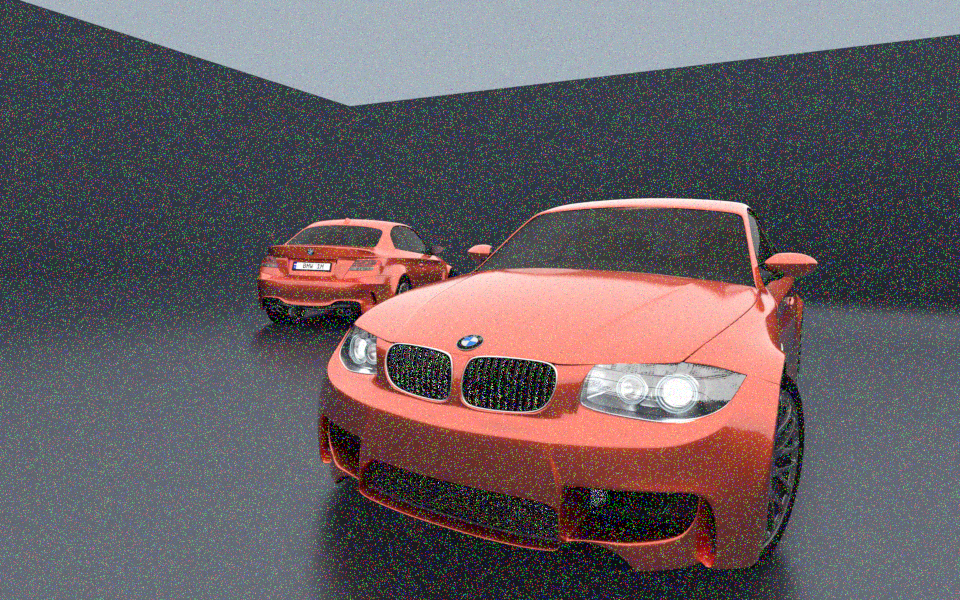}}&
    \subfloat[SSIM = 0.9647\label{fig:SalzReconstBMW}]{\includegraphics[width=0.45\columnwidth]{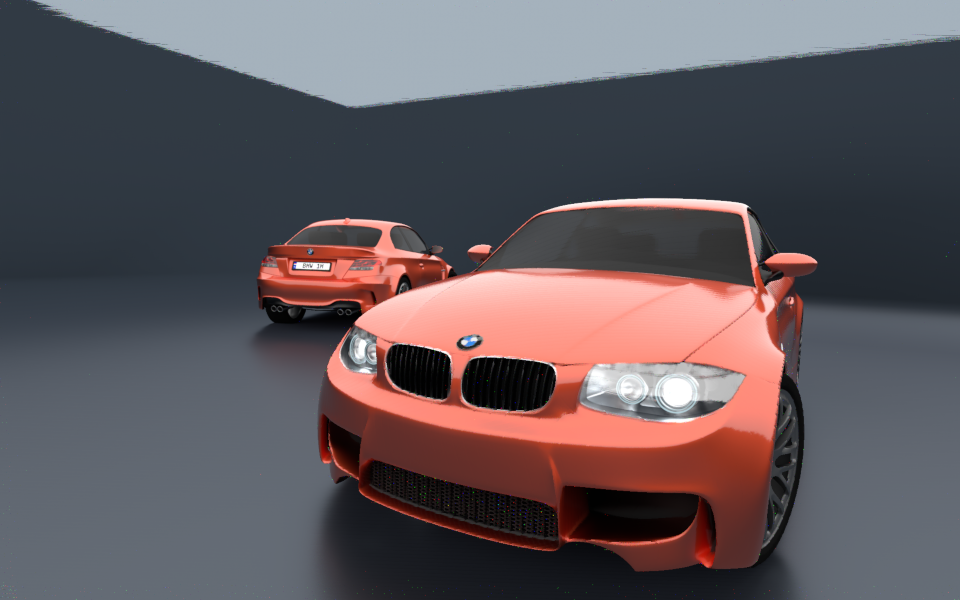}}\\
    {\begin{sideways}{\hspace{0.6cm}Classroom}\end{sideways}}&&
    \subfloat[SSIM = 0.4606\label{fig:GaussNoisyClassroom}]{\includegraphics[width=0.45\columnwidth]{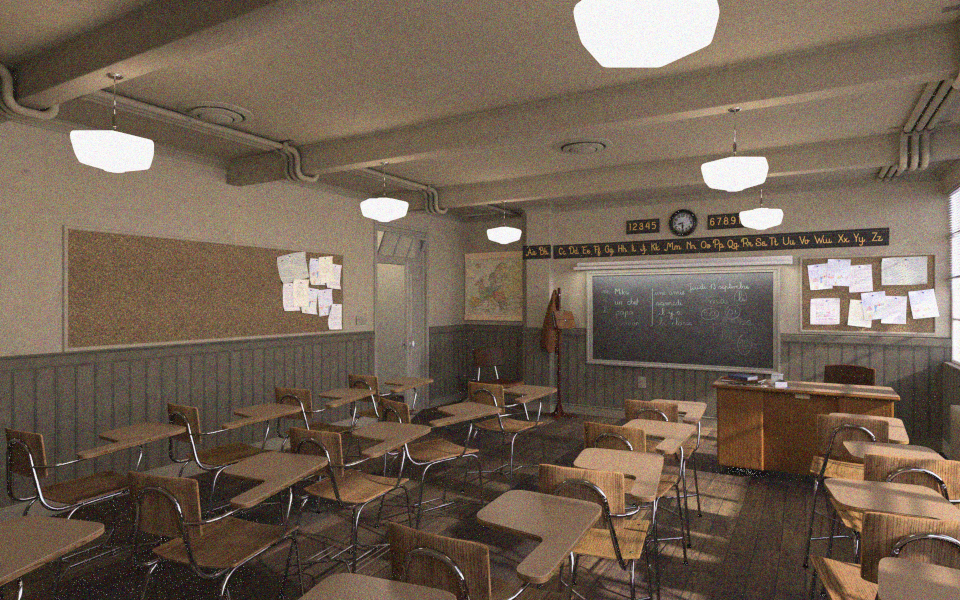}}&
    \subfloat[SSIM = 0.8644\label{fig:GaussReconstClassroom}]{\includegraphics[width=0.45\columnwidth]{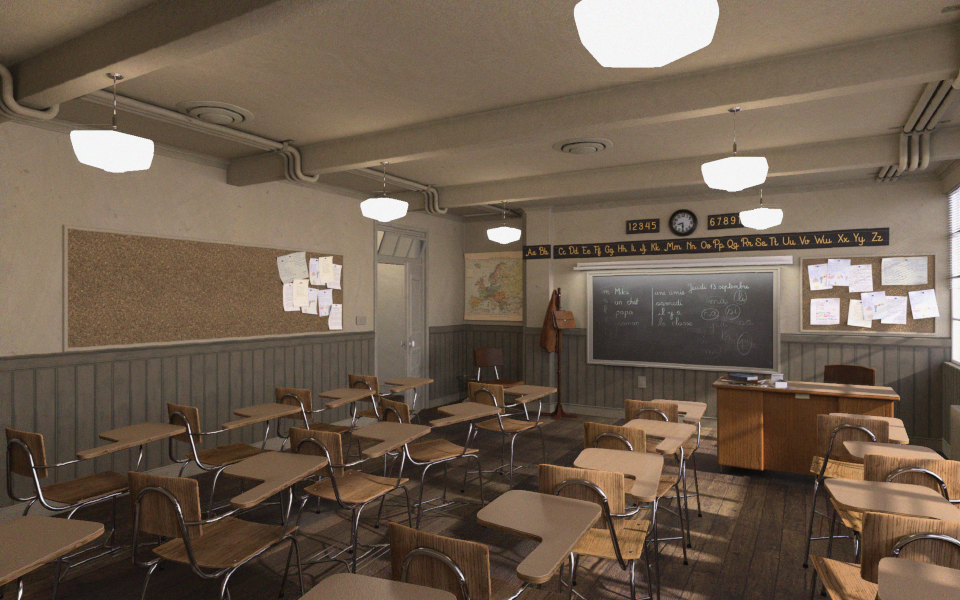}}&
    \subfloat[SSIM = 0.4136\label{fig:SalzNoisyClassroom}]{\includegraphics[width=0.45\columnwidth]{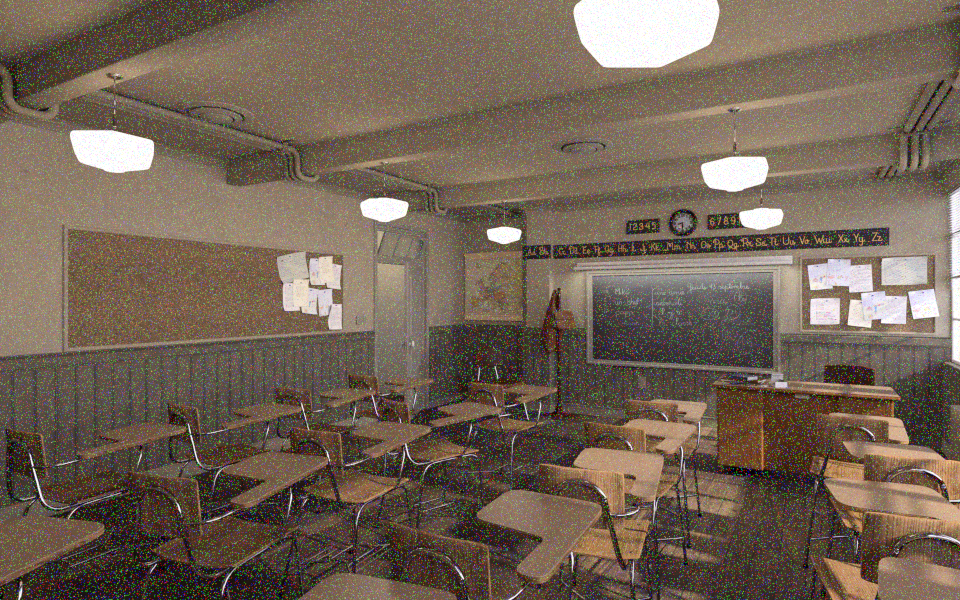}}&
    \subfloat[SSIM = 0.9609\label{fig:SalzReconstClassroom}]{\includegraphics[width=0.45\columnwidth]{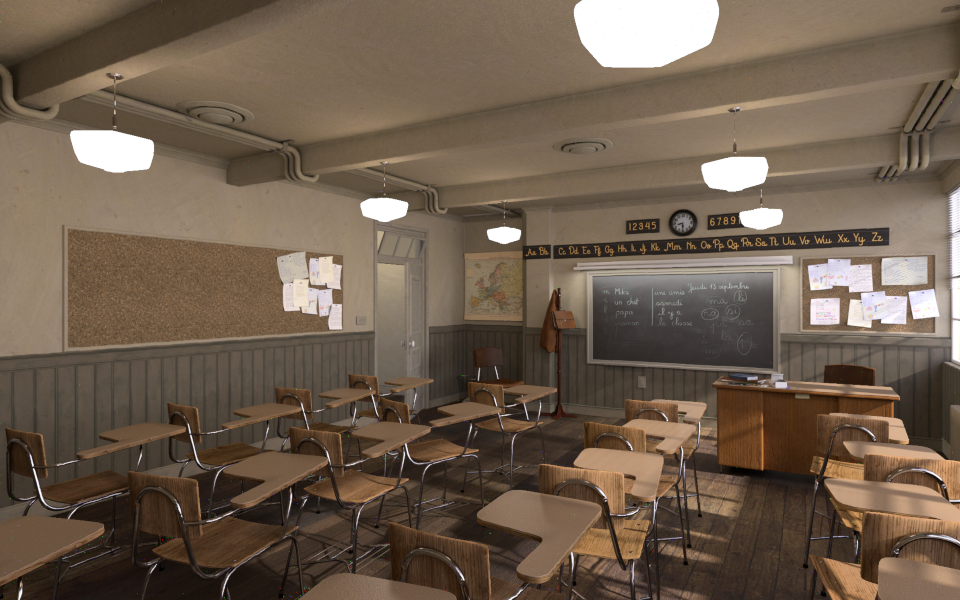}}\\
    {\begin{sideways}{\hspace{0.6cm}Wanderer}\end{sideways}}&&
    \subfloat[SSIM = 0.4350\label{fig:GaussNoisyWanderer}]{\includegraphics[width=0.45\columnwidth]{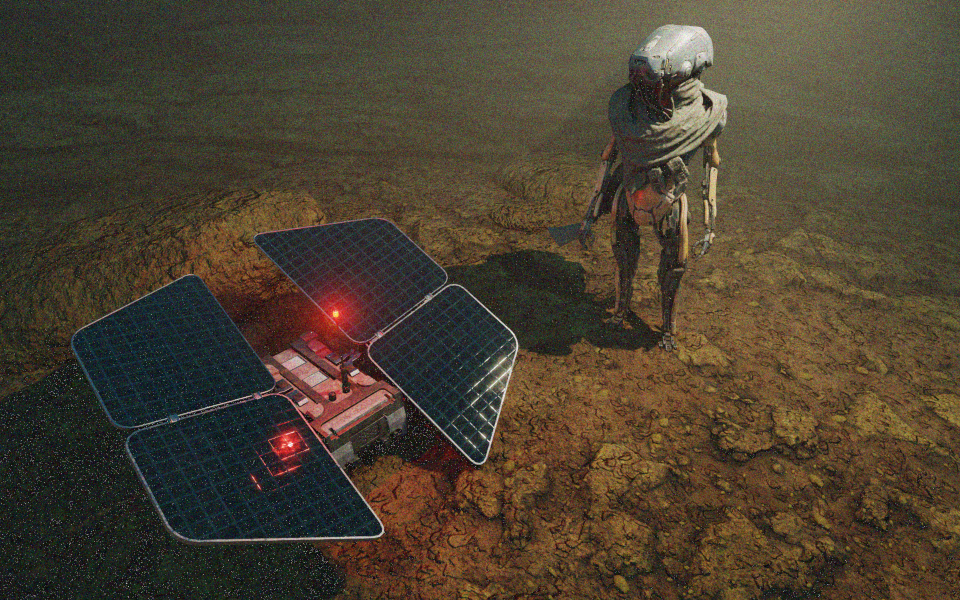}}&
    \subfloat[SSIM = 0.8656\label{fig:GaussReconstWanderer}]{\includegraphics[width=0.45\columnwidth]{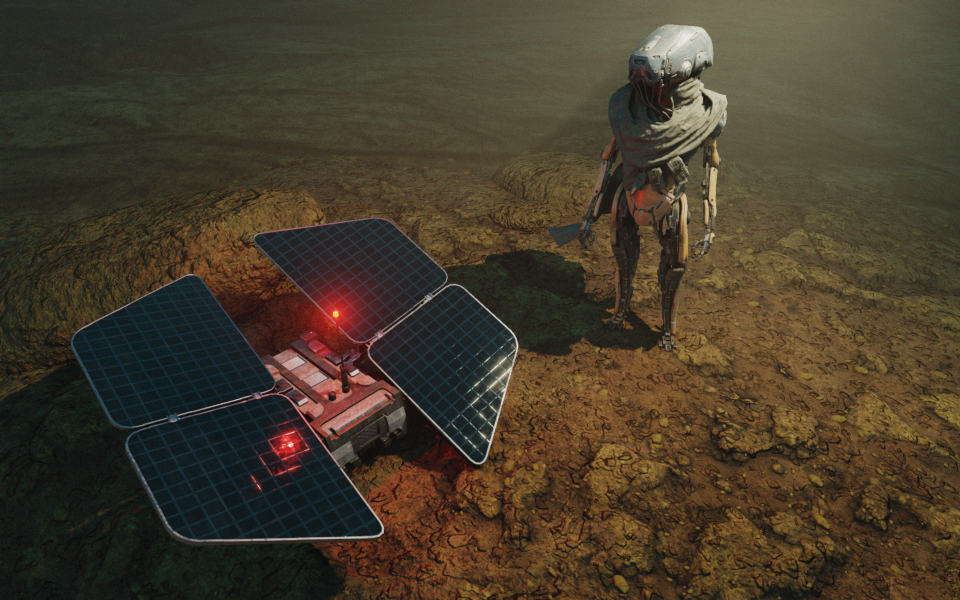}}&
    \subfloat[SSIM = 0.3781\label{fig:SalzNoisyWanderer}]{\includegraphics[width=0.45\columnwidth]{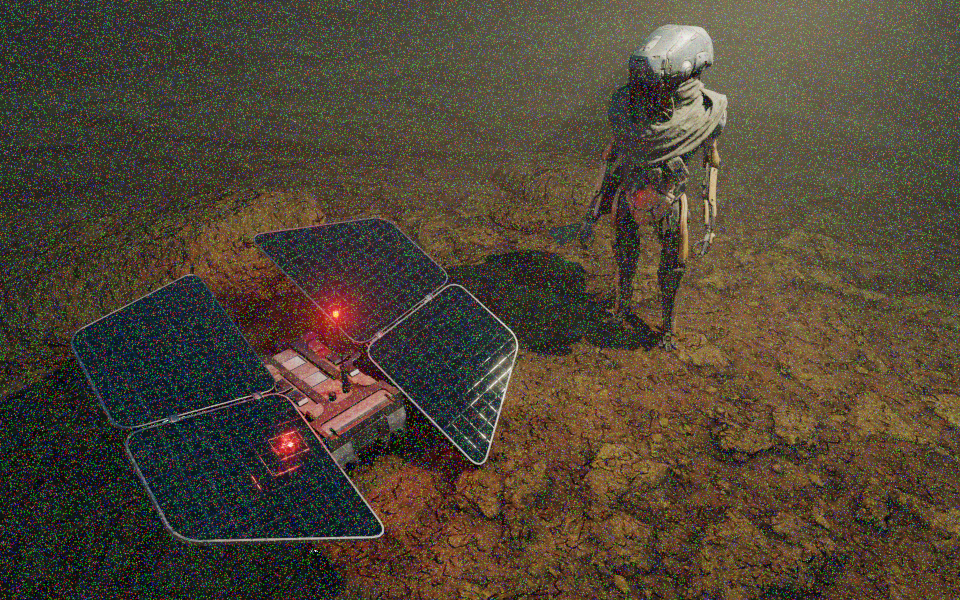}}&
    \subfloat[SSIM = 0.9821\label{fig:SalzReconstWanderer}]{\includegraphics[width=0.45\columnwidth]{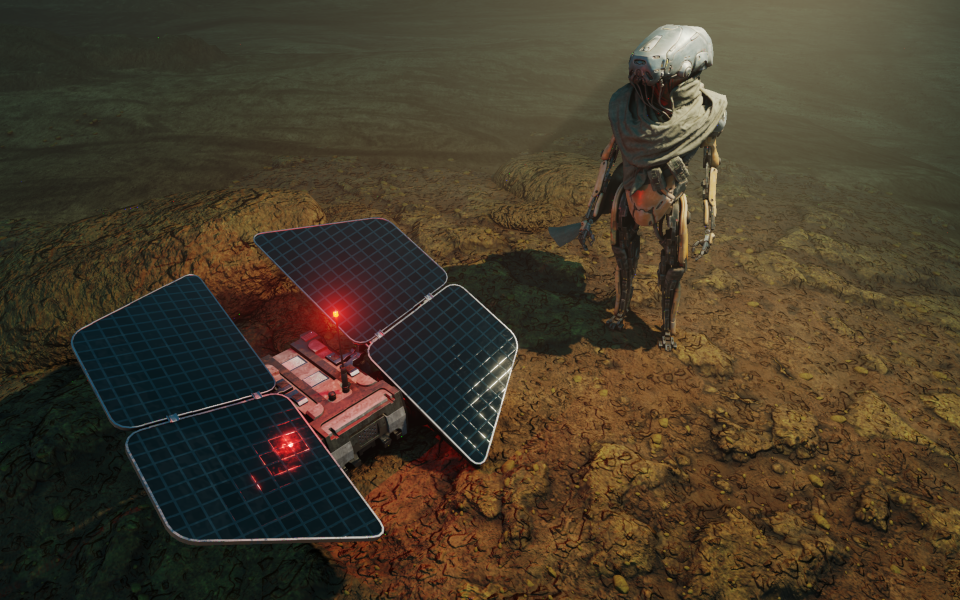}}\\
    \end{tabular}
    \caption{Noisy and filtered LF at viewpoint (a, b) = (2, 2).}
    \label{fig:NoisyVsReconst}%
\end{figure*}
\section{Experimental Analysis\label{sec:Results}}

The semantic representation provided by Fristograms enables us to analyze the distribution of rays and calculate rays for a given froxel. One of the example application endorsing the effectiveness of Fristograms is the substantial reduction of rays used to synthesize an image for a given viewpoint. For example, we rendered the `Classroom' demo scene
using the Blender addon
that also generates the depth maps~\cite{blenderaddon,honauer2016benchmark,blenderdemofiles}. Considering a 4 $\times$ 4 LF camera array (a sub-array of our physical 8 $\times$ 8 array) with each camera having a resolution of 1920 $\times$ 1200 pixels, the total number of rays can be computed as 1920 $\times$ 1200 $\times$ 4 $\times$ 4 $=$ 36,864,000. These rays are found to be distributed among 3,185,991 froxels. Note that, the distribution of rays is specific to the captured scene. \Cref{fig:cdf} depicts the cumulative distribution function of the ray count for the `Classroom' scene, with empty froxels omitted. Calculating a single ray for these froxels using a median filter will therefore reduce the ray count from 36,864,000 to 3,185,991 which is a reduction by a factor of 11.5 for this scene. The maximum achievable reduction in this scenario would be a factor of 16 if every visible froxel is covered by every camera. Due to occlusions in the scene, where some cameras can see behind obstacles, the effectively achievable factor is lower than the maximum. \Cref{fig:OrigVsReconst} shows the comparison between the original LF and the image synthesized using the reduced rays at the same view point\footnote{All the images shown in the paper are provided with full resolution in the code repository}. The small amount of visible artifacts are related to inconsistencies between the images and the corresponding depth maps. Their analysis and removal is out of the scope of this paper.

Fristograms also provide information about the color distribution in a froxel that can be used to perform noise reduction. The effect of noise on the rays in a froxel is already shown in \Cref{fig:NoiseAnalysis}. As a simple demonstration of the capabilities of fristograms we used this information to perform noise reduction by using a mean or median filter. As shown in \Cref{fig:NoisyVsReconst}, we add the same AWGN noise as used above to the LF (\Cref{fig:GaussNoisyBMW,fig:GaussNoisyClassroom,fig:GaussNoisyWanderer}) and filter the froxel data using a mean filter. The improvement in the quality is significant and visually evident (\Cref{fig:GaussReconstBMW,fig:GaussReconstClassroom,fig:GaussReconstWanderer}). The SSIM score for the synthesized viewpoint also confirms the reduction in noise. In contrast to traditional image filtering techniques, filtering the froxel data always keeps edges and textures. For froxels with sufficient samples, the improvements are comparable to traditional techniques which take the data from neighboring pixels into account. As shown in \Cref{fig:NoisyVsReconst} we also simulated salt-and-pepper noise with a noise density of 5$\%$. \Cref{fig:SalzNoisyBMW,fig:SalzNoisyClassroom,fig:SalzNoisyWanderer} show the noisy LF images. We then performed the same ray allocation and performed a median filtering on the froxel data. The synthesized images (\Cref{fig:SalzReconstBMW,fig:SalzReconstClassroom,fig:SalzReconstWanderer}) show a significant improvement in visual quality and is also confirmed by the SSIM score. While the chosen filters are appropriately chosen for the given noise type, the capabilities of Fristrograms are applicable on a more general level.



\section{Conclusion\label{sec:Conclusion}}

In this paper we introduce a novel representation for LFs, the Fristogram. The scene frustum is discretized into froxels, which cover a configurable amount of pixels in width, height and disparity between neighboring cameras. Assigning all rays of the LF to the froxels from which they originate creates Fristograms. They enable a simple differentiation between Lambertian and non-Lambertian materials as well as new filtering techniques not requiring information from surrounding pixels. Additional benefits include the possibility to detect other material properties, employ deep learning techniques to classify froxels, perform material specific volumetric filtering and implement advanced noise reduction schemes. With these simple to implement examples, we show the power of the representation of LFs as Fristograms, which comes from the inherent combination of ray data with the scene and camera geometry.


\bibliographystyle{IEEEbib}
\bibliography{arxiv}

\end{document}